\documentclass[11pt,journal, letterpaper]{IEEEtran}
\IEEEoverridecommandlockouts
\usepackage{amsmath,amssymb,amsfonts}
\usepackage{algorithmic}
\usepackage{graphicx}
\usepackage{comment}
\usepackage{textcomp}
\usepackage{xcolor}
\usepackage{cite}

\usepackage[11pt]{moresize}

\usepackage{hyperref}

\usepackage[percent]{overpic}

\usepackage{tabularx}

\usepackage{multicol}
\usepackage{boldline}

\usepackage{caption}% http://ctan.org/pkg/caption
 \usepackage{background}
 \backgroundsetup{angle=0,  
   scale=1, 
   color=black, 
   firstpage=true,
   position=current page.north, 
   hshift=0pt,
   vshift=-20pt,
   contents={\ifnum\value{page}=1 PREPRINT: Accepted for publication at the IEEE Consumer Electronics Magazine (CEM). \else \fi}
 }

\def\BibTeX{{\rm B\kern-.05em{\sc i\kern-.025em b}\kern-.08em
    T\kern-.1667em\lower.7ex\hbox{E}\kern-.125emX}}
 
\begin{document}

\title{\mbox{Approximate LSTMs for Time-Constrained Inference}: Enabling Fast Reaction in Self-Driving Cars}

%\author{\IEEEauthorblockN{Alexandros Kouris, Stylianos I. Venieris, Michail Rizakis and Christos-Savvas Bouganis}\\
%\IEEEauthorblockA{\textit{Electrical \& Electronic Engineering Department}, 
%\textit{Imperial College London, UK}\\
%\{a.kouris16,stylianos.venieris10,michail.rizakis14,christos-savvas.bouganis\}@imperial.ac.uk}
%\ \vspace{-7mm}
%}

\author{\IEEEauthorblockN{Alexandros Kouris$^\dagger$, Stylianos I. Venieris$^\ddagger$, Michail Rizakis$^\dagger$ and Christos-Savvas Bouganis$^\dagger$}\\
\IEEEauthorblockA{ $^\dagger$\textit{Electrical \& Electronic Engineering Department}, 
\textit{Imperial College London, UK}\\
email: \{a.kouris16,michail.rizakis14,christos-savvas.bouganis\}@imperial.ac.uk}\\
\IEEEauthorblockA{$^\ddagger$\textit{Samsung AI Center (SAIC)}, 
\textit{Cambridge, UK} - 
email: s.venieris@samsung.com}
\ \vspace{-7mm}
}

\maketitle

\begin{abstract}
The need to recognise long-term dependencies in sequential data such as video streams has made Long Short-Term Memory (LSTM) networks a prominent Artificial Intelligence model for many emerging applications. However, the high computational and memory demands of LSTMs introduce challenges in their deployment on latency-critical systems such as self-driving cars which are equipped with limited computational resources on-board. In this paper we introduce a progressive inference computing scheme that combines model pruning and computation restructuring leading to the best possible approximation of the result given the available latency budget of the target application.
The proposed methodology enables mission-critical systems to make informed decisions even in early stages of the computation, based on approximate LSTM inference, meeting their specifications on safety and robustness. Our experiments on a state-of-the-art driving model for autonomous vehicle navigation demonstrate that the proposed %approximate computing scheme %implemented on a FPGA 
approach can yield outputs with similar quality of result compared to a faithful LSTM baseline, up to 415$\times$ faster (198$\times$ on average,\mbox{ 76$\times$ geo. mean).}
%Our experiments demonstrate that using the proposed methodology, mission-critical systems responsible for autonomous navigation and collision avoidance are able to make informed decisions based on approximate calculations within the available time budget, meeting their specifications on safety and robustness.
\end{abstract}

%\begin{IEEEkeywords}
%component, formatting, style, styling, insert
%\end{IEEEkeywords}

\vspace{-0.4cm}
\section{Introduction}
\label{sec:intro}

Recurrent neural networks (RNNs) are a family of \mbox{machine} learning models with the ability to recognise patterns in sequential and temporal data. In the past decade, long short-term memory (LSTM) networks \cite{lstm1997} have emerged as the dominant RNN by setting the state-of-the-art record in various AI tasks, such as machine translation and video understanding.
%spanning from human trajectory prediction \cite{Alahi_2016_CVPR} to video understanding \cite{otte2016recurrent}. 
Among the various LSTM-enabled applications,  time-constrained mission-critical systems \cite{ce1} are rapidly becoming an ubiquitous scenario. In this setting, AI agents are equipped with LSTM-based mechanisms of sensing, perceiving and, eventually, acting\cite{ce6}. In such scenarios, making the most informed decision under a limited time budget is of vital importance in order to ensure the robust, safe and successful operation of the system within complex and uncertain environments \cite{ce2}. 

\begin{figure}
  \vspace{-5mm}
  \centering
  \includegraphics[trim={5mm 10mm 25mm 10mm}, clip,width=0.5\textwidth]{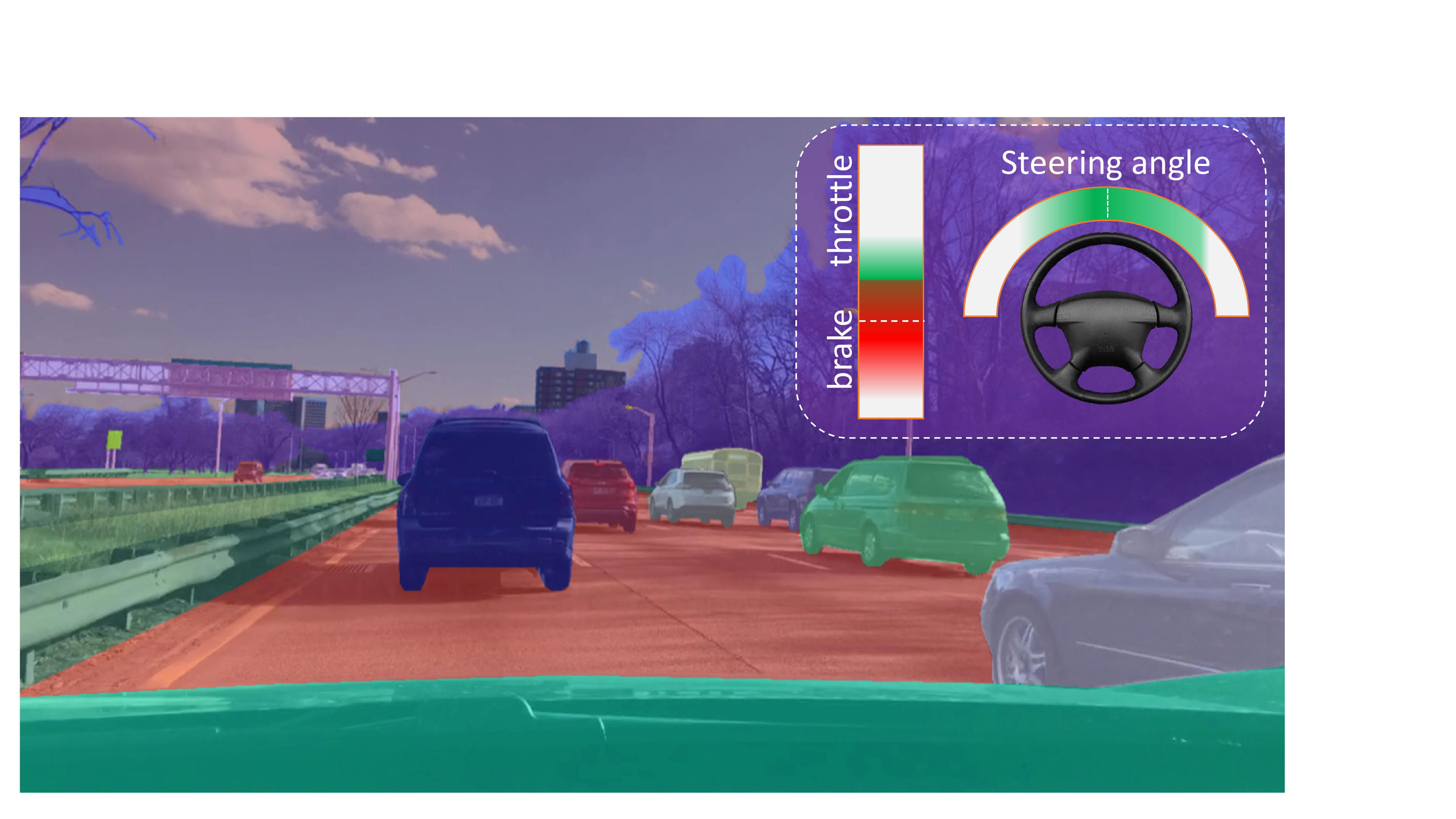}
  \caption{Throttle/brake and steering angle prediction for autonomous driving with an LSTM model (trained on the dataset of \cite{yu2018bdd100k}), relying on visual inputs. \\ \vspace{-2mm} { %\scriptsize
  \ssmall \textbf{Video \& Webpage:}}{ %\tiny
  \ssmall \href{https://www.imperial.ac.uk/intelligent-digital-systems/approx-lstms/}{ \color{blue} www.imperial.ac.uk/intelligent-digital-systems/approx-lstms/}}
  }
  \vspace{-5mm}
  \label{fig:autonDriving}
\end{figure}

Fig. \ref{fig:autonDriving} depicts an example of such a latency-critical system. In this case, a driverless car navigates autonomously in an urban environment under the control of an LSTM that predicts the desired throttle/brake position and steering angle based on the input video sequence. With human driver reaction time ranging between 0.7 and 3 seconds (varying with situation and individual person) \cite{mcgehee2000driver}, autonomous driving systems target a relevant low-latency envelope to take action from the moment an event occurs on the road, in order to preserve the ability of achieving comparable reliability with humans. In this respect, extracting the best possible approximation of the desired action to be commanded within the real-time latency constraints is preferred from a more accurate decision later in time.

\begin{figure}[t]
%   \vspace{-2mm}
  \centering
  \includegraphics[trim={7cm 7cm 7cm 4.5cm},clip,width=0.41\textwidth]{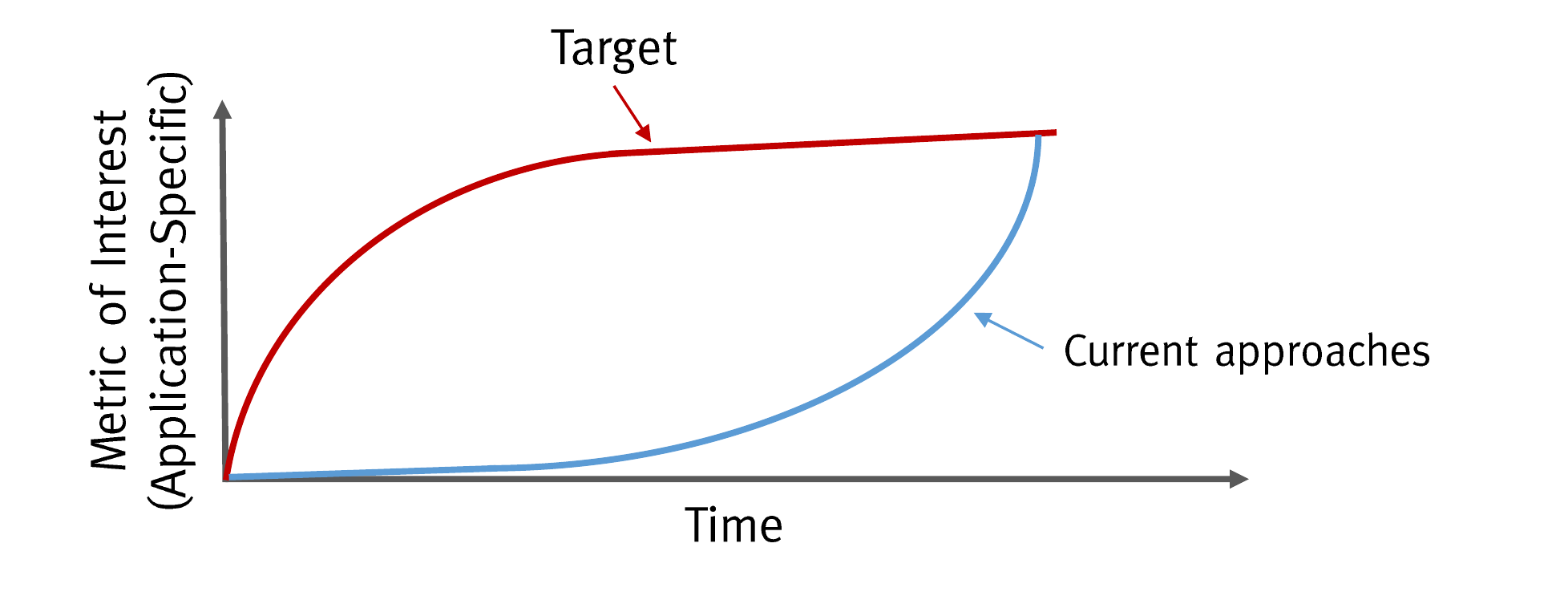}
  \vspace{-0.4cm}
  \caption{ The concept of progressive inference: Conventional and target behaviour of time-constrained AI systems. The y-axis metric reflects the application-level accuracy (higher-is-better). }
  \label{fig:behaviours}
  \vspace{-0.6cm}
\end{figure}

From a technical viewpoint, performing the most informed action under a time budget reduces to the problem of obtaining the highest quality output from an LSTM given a constraint in computation time. Current methods of deploying LSTMs follow the behaviour depicted in \mbox{Fig. \ref{fig:behaviours}}. Conventional implementations \cite{Guan_2017, Chang_2017
%, Zhang_2017}
} require the whole inference computation to finish in order to obtain meaningful information from the LSTM and thus prolong the sensing-to-action loop with potentially catastrophic effects. Instead, the stringent latency deadlines of real-life systems call for \textit{progressive inference} designs that can provide the best possible estimate of their final output for a given time budget and improve on it as more time budget becomes available (Fig. \ref{fig:behaviours}). This property would enable the agent to exploit the maximum possible amount of information that is available in the current input and effectively optimise its overall operation.

From a workload perspective, LSTMs are challenging by being memory-bound. This property means that the performance of brute-force implementations is limited by the available memory bandwidth of the platform, rather than by the available computational power.  Furthermore, the excessive memory accesses and the inefficient use of computational resources when executing LSTMs on conventional platforms leads to substantial power inefficiencies which are critical for battery-constrained settings. {To attack this issue, recent works deviated from general-purpose computing platforms and adopted a \textit{model-hardware co-design} approach for the generation of custom  FPGA-based  hardware architectures \cite{ce3}. Field-programmable gate arrays (FPGAs) typically consist of one or more processors and a reconfigurable fabric. The processor is responsible for executing non-critical code and coordinates the operation of the overall system. The reconfigurable fabric can be customised at the hardware level, allowing the on-chip  computational and memory resource allocation to be optimised to match the particular workload and the performance needs of the target application and its underlying implementation.

Enabled by the customisation and flexibility of FPGAs, the works below propose different approximation techniques, focusing on model compression \cite{Han_2017}, quantisation \cite{Wang_2017
%, Rybalkin_2018}
} and pruning \cite{ Zhang_2017
%, Wang_2017
}, together with an associated FPGA-based hardware accelerator, tailored to the computational needs of the model and its approximate computing scheme, to match the computational demands of LSTMs.} Despite the effectiveness of these methods, their application requires a \textit{retraining step}, which allows the refinement of the model in order to compensate for any approximation losses in the model's accuracy. For the retraining step to be feasible, availability of the training set is required, which is not a realistic assumption in privacy-aware applications \cite{ce9}, as in the case of large-scale datasets collected by commercial companies that remain proprietary, %while leading to the development of state-of-the-art AI models
or medical-oriented institutions that are prevented by confidentiality regulations from sharing their clinical datasets, making privacy-preserving AI techniques increasingly relevant \cite{shokri2015privacy,kouris18cascade,wainwright2012privacy}.

In this context, we propose a novel methodology for the high-performance deployment of LSTMs in time-constrained applications, which is also complementary to the existing approaches. The proposed approximate computing scheme is implemented on custom hardware, also exploiting the customisation and flexibility of FPGAs.
%An FPGA typically consists of one or more processors and a reconfigurable fabric. The processor is responsible for executing non-critical code and coordinates the operation of the overall system. The reconfigurable fabric can be customised at the hardware level and in this way the on-chip computational and memory resource allocation %, as well as the input-output (I/O) subsystem, 
%can be optimised to match the particular workload and the performance needs of the target application.
%The reconfigurable fabric of an FPGA can be customised at the hardware level, allowing the on-chip computational and memory resource allocation to be optimised to match the particular workload and the performance needs of the target application.
The goal is to generate an optimised hardware mapping of a given LSTM on a target FPGA, tailored to the available time budget and error tolerance. To meet the needs of this task, an iterative scheme is introduced that exploits the resilience of the target application to approximations in order to relax the computational and memory requirements of the given LSTM, and executes the model under time constraints, with increasing accuracy as a function of the time budget. % by means of refinement iterations.

In this work, we showcase a significantly improved computation-time, accuracy and power trade-off presented by our progressive inference scheme that effectively reduces the computational workload of a given LSTM model to meet the desired quality of result, compared to a baseline implementation of the same model, while both designs are exploiting the customisation capabilities of an FPGA.  The experimental evaluation of the proposed approach is conducted on a state-of-the-art driving model for autonomous vehicles. Self-driving cars, being tightly coupled with the recent developments in Consumer Electronics \cite{ce4} \cite{ce8}, form a representative example of a system with tight computation time budget to make mission critical decisions, while being also constrained in a limited computational resource environment. At the same time, autonomous driving is emerging alongside with the revolution of electric vehicles, imposing a low-power envelope for the deployment of increasingly compute-hungry models \cite{ce7}. This makes special-purpose FPGA-based hardware architectures the most prominent solution, offering high computational efficiency for deployment on resource- and power-constrained environments.

\vspace{-2mm}
\section{Learning long-term patterns with LSTMs}
\label{sec:background}
\begin{figure}[t]
  \vspace{-6.5mm}
  \centering
  \includegraphics[trim={4cm 5.5cm 10cm 4.5cm},clip,width=0.5\textwidth]{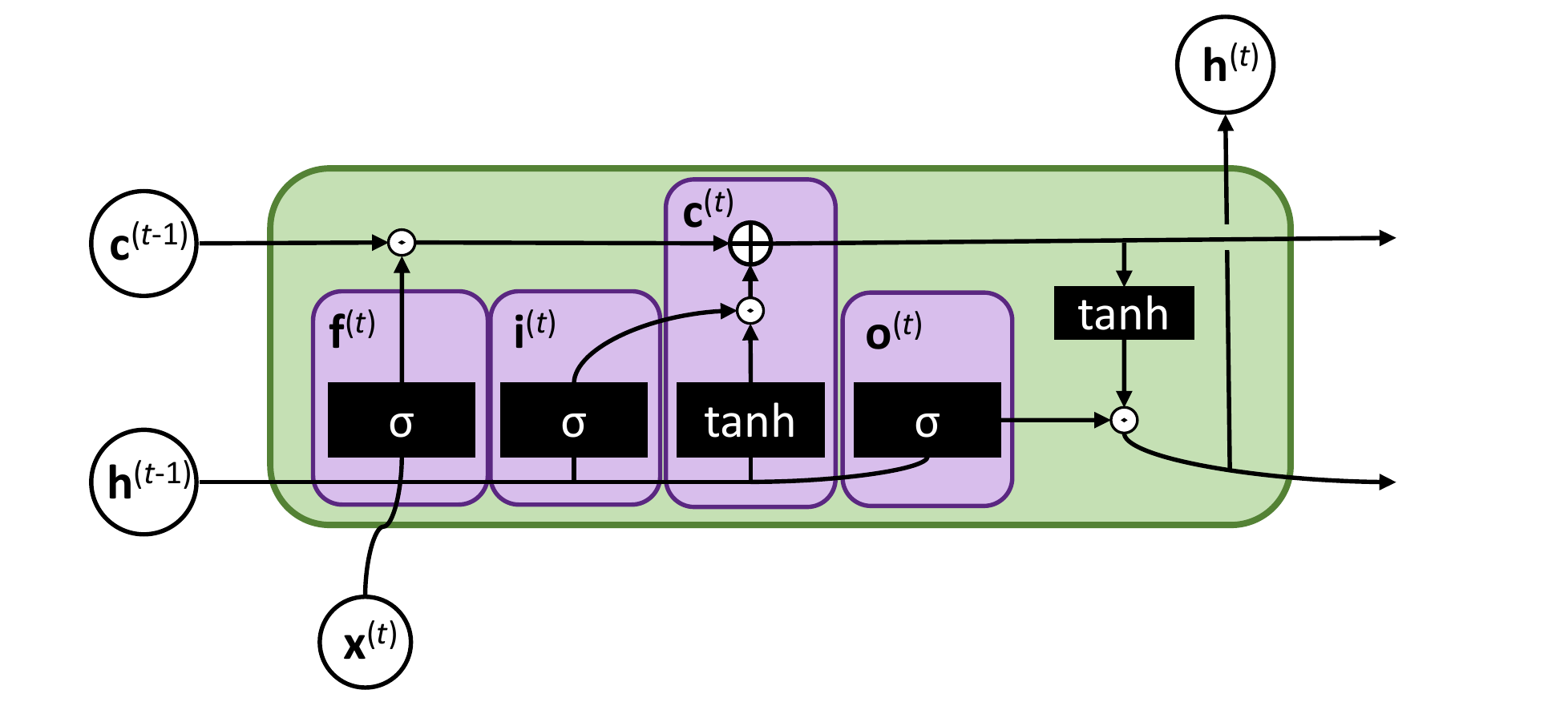}
      \put (-160,16) {\footnotesize $\textbf{g}^{(t)} = \sigma(\textbf{W}_\text{x} \textbf{x}^{(t)} + \textbf{W}_\text{h} \textbf{h}^{(t-1)})$}
    \put (-160,3) {\footnotesize $\textbf{y}^{(t)} = \textbf{y}^{(t-1)} \odot \textbf{g}^{(t)}$ }
  \caption{Structure of an LSTM model.  $\mathbf{g}^{(t)}$ represents each of the LSTM gates ($\mathbf{f}^{(t)},\mathbf{i}^{(t)},\mathbf{c}^{(t)},\mathbf{o}^{(t)}$), while $\odot$ denotes the element-wise multiplication between two vectors defined as $(\textbf{a}\odot \textbf{b})_i = \textbf{a}_i \textbf{b}_i$.  }
  \label{fig:lstm}
  \vspace{-0.5cm}
\end{figure}

LSTMs are specialised RNNs with enhancements that enable the learning of long-term dependencies. The key feature of an LSTM is a set of units named \textit{gates} which control its behaviour at run time. Fig. \ref{fig:lstm} depicts the structure of an LSTM. The core element of LSTMs is the cell state $\textbf{c}$, shown along the horizontal line at the top of the diagram. At each time step $t$, the LSTM removes or adds information to the cell state via its gate modules. Computationally, a gate receives as inputs the new input sample $\textbf{x}^{(t)}$ and the previous output $\textbf{h}^{(t-1)}$ and performs a matrix-vector multiplication with the weight matrices $\textbf{W}_{\text{x}}$ and $\textbf{W}_{\text{h}}$, as described  in Fig. \ref{fig:lstm}. % on the first line of \mbox{Eq. (\ref{equ:gate})}. 
The elements of the weight matrices are learned during the training stage of the target application and remain fixed throughout the inference stage that takes place upon deployment.

Next, the resulted vector of the matrix-vector multiplication is passed through a nonlinear function, such as a sigmoid $\sigma(\cdot)$, to form $\textbf{g}^{(t)}$. The nonlinear function operates in an element-by-element fashion and outputs a vector with values between 0 and 1, capturing how much of each element should be kept. A value of 0 represents total forgetting of information, 1 represents total propagation and intermediate values dictate what fraction of the information should be kept. In this manner, by multiplying element-by-element another vector $\textbf{y}^{(t-1)}$ with the output of the nonlinear function,  a new vector $\textbf{y}^{(t)}$ is produced which is a filtered version of its previous state  (Fig. \ref{fig:lstm}).%:
\begin{comment}
\begin{equation}
\label{equ:gate}
\begin{aligned}
    \textbf{g}^{(t)} &= \sigma(\textbf{W}_\text{x} \textbf{x}^{(t)} + \textbf{W}_\text{h} \textbf{h}^{(t-1)})\\
    \textbf{y}^{(t)} &= \textbf{y}^{(t-1)} \odot \textbf{g}^{(t)} 
\end{aligned}
\end{equation}
where $\odot$ denotes the element-wise multiplication between two vectors defined as $(\textbf{a}\odot \textbf{b})_i = \textbf{a}_i \textbf{b}_i$. 
\end{comment}

An LSTM consists of four gates. Starting from the left of the diagram in Fig. \ref{fig:lstm}, the \textit{forget} gate $\textbf{f}^{(t)}$ determines the amount of information that will be forgotten from the previous cell state $\textbf{c}^{(t-1)}$. Next, the \textit{input} gate $\textbf{i}^{(t)}$ and the \textit{cell} gate determine the new information to be stored in the new cell state $\textbf{c}^{(t)}$. The cell gate employs tanh for its nonlinear function and creates a vector of new candidate values for the new cell state, while the input gate controls which values of the current cell state will be updated. At this point, the new cell state $\textbf{c}^{(t)}$ has been formed. The final step involves the calculation of the new output vector $\textbf{h}^{(t)}$, which is a filtered version of the cell state. This is generated by passing the cell state through a tanh nonlinearity and multiplying the result with the output of the \textit{output} gate $\textbf{o}^{(t)}$ in order to update only parts of the cell state.

\vspace{-3mm}
\section{Approximate Computing For LSTMs}
\label{sec:approx}
At the core of an LSTM's workload lie the linear algebra operation of matrix-vector multiplication, shown on the first line in Fig. \ref{fig:lstm}, which takes place in each of the four gates. Neural networks have been extensively studied to have redundancy in terms of their trained parameters \cite{Denil_2013}. This property allows the restructuring of the computations of LSTM gates in such a manner that enables us to extract the maximum information at any time instant. In this respect, we propose an approximate computing scheme that enables the tuning of the quality of result (QoR) in exchange for an increase in performance. The proposed approach exploits the statistical redundancy of LSTMs by acting at two levels: \mbox{(i) approximating} weight matrices with a low-rank Singular-Value Decomposition (SVD) and \mbox{(ii) pruning} the network by sparsifying the weight matrices based on an importance criterion of their elements. These techniques enable us to restructure the computations of an LSTM and design a computing system that performs the most information-carrying computations first in order to obtain the peak QoR given a time budget.

\textbf{Information-maximising approximation.}
Each LSTM gate consists of two weight matrices corresponding to the current input and previous output respectively. In our scheme, we first concatenate the two weight matrices and the input and output vectors to obtain a single augmented matrix and vector respectively for each gate as $\textbf{W} = [\textbf{W}_{\text{x}} \textbf{W}_{\text{h}}] \in \mathbb{R}^{R\times C}$ and \mbox{$\Tilde{\textbf{x}}^{(t)} = \left[ \textbf{x}^{(t)\top} \textbf{h}^{(t-1)\top} \right]^{\top} \in \mathbb{R}^{C\times 1}$}. As a next step, we substitute the augmented weight matrix with a low-rank approximation that reduces the computation and memory footprint cost while minimising the information loss. These properties are satisfied by the rank-1 approximation of each weight matrix based on the SVD. This approach enables us to approximate the weight matrix as the outer product of two vectors (the singular vectors) followed by an elementwise multiplication with a constant number (the singular value). For the i-th gate, the rank-1 approximate weight matrix is given by $\widetilde{\textbf{W}}_i = \sigma^i_1 \textbf{u}^i_1 \textbf{v}^{i\top}_1$. With respect to computational cost, the original matrix vector multiplication $\widetilde{\textbf{W}}_i \Tilde{\textbf{x}}^{(t)}$ is replaced by a dot product followed by an elementwise multiplication between a vector and a constant number, \textit{i.e.} $\sigma_1^i \textbf{u}_1^i (\textbf{v}_1^{i\top} \Tilde{\textbf{x}}^{(t)})$, leading to a significant reduction on both the number of operations and the memory footprint of the weight matrix, while retaining the highest amount of information that a rank-1 approximation can have.

\textbf{Pruning by means of network sparsification.}
The second level of approximation on the LSTM comprises the structured pruning of the weight matrices at each gate. Pruning can interpreted as a type of sparsity in which individual weights are masked as zeros. In our structured pruning scheme, we limit sparsity to the structure of rows of the weight matrices. This selection of granularity allows us to always obtain an approximate value for each element of the resulted output vector, instead of having zeroed values at the output vector that carry no information. Individual weight values are set to zero by means of a magnitude-based criterion which determines the importance of a weight using its absolute value.  Overall, the pruning scheme preserves the NZ elements with the highest absolute value on each row of each weight matrix. The value of NZ is tuned to provide the highest possible application-level accuracy, considering the user-specified latency budget.
%leading to NZ non-zero elements per row.

\begin{comment}
This can be expressed as:
\vspace{-2mm}
\begin{equation}
\vspace{-2mm}
    \label{equ:pruning}
    \Tilde{\textbf{w}}^{\text{pruned}}_j = \text{prune}(\Tilde{\textbf{w}}_j, \text{NZ}), ~~ \text{for all rows $j$ of matrix $\widetilde{\textbf{W}}$}
\end{equation}
where $\Tilde{\textbf{w}}_j$ is the j-th row vector of matrix $\widetilde{\textbf{W}}$ and NZ determines the desired sparsity level of the resulting vector $\Tilde{\textbf{w}}^{\text{pruned}}_j$ and is tuned to provide the highest possible application-level accuracy, considering the user-specified latency budget.
\end{comment}

\textbf{Hybrid compression and pruning.}
To obtain a refinement mechanism that allows us to increase the quality of result as a function of time while leveraging the advantages of both aforementioned techniques, we combine them in a hybrid iterative approximation method given by Eq. (\ref{equ:hybrid_approx}). The iterative nature of the hybrid method involves the refinement of the computed output over a number of iterations, with each refinement step involving the addition of a low-rank approximation of a correction factor (residual) together with its pruning.
\vspace{-2mm}
\begin{equation}
\vspace{-2mm}
    \label{equ:hybrid_approx}
        \Tilde{\textbf{y}}_i = \sum_{n=1}^{N_{\text{steps}}} \{\underbrace{ \sigma^{i(n)}_1 \textbf{u}^{i(n)}_1 ( \overbrace{\text{prune}({\textbf{v}}^{i(n)}_1, \text{NZ})^\top}^{\text{pruning}} \Tilde{\textbf{x}}^{(t)} )}_{\text{refinement step}} \} 
    % \Tilde{\textbf{y}}_i = \sum_{n=1}^{N_{\text{steps}}} \{\underbrace{ ( \overbrace{\text{prune}(\Tilde{\textbf{w}}^{(n)}_i, \text{NZ})^\top}^{\text{pruning}} \Tilde{\textbf{x}}^{(t)} )}_{\text{refinement step}} \} 
\end{equation}
With this scheme, the final approximate output vector is formed after applying $N_{\text{steps}}$ refinement steps. The weight matrices of each LSTM gate are approximated by $N_{\text{steps}}$ %singular 
vector pairs. At the n-th refinement iteration, the %singular 
value $\sigma_1^{i(n)}$ and vectors $\textbf{u}_1^{i(n)}$ and $\textbf{v}_1^{i(n)}$ capture the rank-1 approximation of a correction factor. In this manner, at each refinement step, the current $\textbf{v}_1^{i(n)}$ %singular 
vector is pruned using our pruning scheme, in order to end up with NZ non-zero elements, and then is multiplied with the current augmented input vector, resulting to an non-full \mbox{rank-1} approximation. By utilising the approximation \textit{residual} at each time step \mbox{($\textbf{R}_i^{(n)}=\textbf{W}_i$-$\widetilde{\textbf{W}}_i^{(n-1)}$)} to extract an SVD-based rank-1 correction factor for the progressive refinement of the augmented weight-matrix approximation, the error due to both the SVD and the pruning are considered in contrast to the case of progressively applying higher-rank approximations of the original weight matrix, minimising in this way the information loss \cite{rizakis18approximate}. Hence, the workload of each gate is reduced to $N_{\text{steps}}(2R+2NZ+1)$ operations.

%By applying this process for $N_{\text{steps}}$, the final approximation of the output vector is produced. 

Therefore, in the hybrid method, different combinations of level of pruning and number of refinement steps correspond to different candidate designs with different computation cost and QoR. In this respect, the number of non-zeros (NZ)  and the number of refinements ($N_{\text{steps}}$) form tunable parameters that are optimised by the proposed methodology to meet the time constraints and QoR requirements of the target application.

\vspace{-4mm}
\section{Domain-specific Architecture for LSTMs}%A Custom Hardware Architecture for LSTMs}
\vspace{-0mm}
\label{sec:arch}

%TODO We need to highlight the main features of architecture which are LSTM-specific. These could be: 1) adoption of the dataflow instead of the control-flow operation, i.e. data-driven instead of instruction-based, 2) exploitation of parallelism both between and within the LSTM gates (inter- and intra-gate parallelism), 3) parametrisation so that the scaling of the accelerator can be controlled at compile time (i.e. parameters Tc and Tr), 4) precomputation of singular vectors and values (?) - this might be a detail.

The philosophy behind the proposed architecture is to overcome the limitations of programmable processors by introducing a set of strategies that exploit the properties of LSTMs. These include the adoption of \textit{dataflow processing} to alleviate the overheads of conventional computing platforms, the exploitation of both the \textit{inter-gate} and \textit{intra-gate parallelism} of LSTMs to boost performance and the compile-time \textit{tunable scaling} of the architecture to match the available resources and the response-time demands of the target application. 

\textbf{Dataflow processing.}
In contrast with the control-flow paradigm of general-purpose computers where individual instructions are scheduled for execution, we adopt a data-driven dataflow architecture. In this scheme, the availability of input samples triggers the LSTM processing to be performed on them without the need for explicit control and synchronisation between computation units. From a hardware perspective, this approach allows us to remove any generic instruction-handling hardware logic and repurpose the resources of the FPGA chip specifically for LSTMs. In this way, the architecture avoids the time, resource and power overhead of off-the-shelf platforms and boosts the attainable performance by dedicating more hardware resources for computation.

% where the availability of input samples triggers the LSTM processing to be performed on them. The proposed architecture exploits the inherent parallelism both between and within the four gates of the LSTM and is parametrised in order to allow the compile-time tuning of the performance-resource cost trade-off. 

\begin{figure}[t]
%   \vspace{-2mm}
%  \centering  \includegraphics[trim={10mm 135mm 280mm 80mm },clip,width=0.5\textwidth]{arch_v6.pdf}
   \centering  \includegraphics[trim={10mm 135mm 280mm 80mm },clip,width=0.5\textwidth]{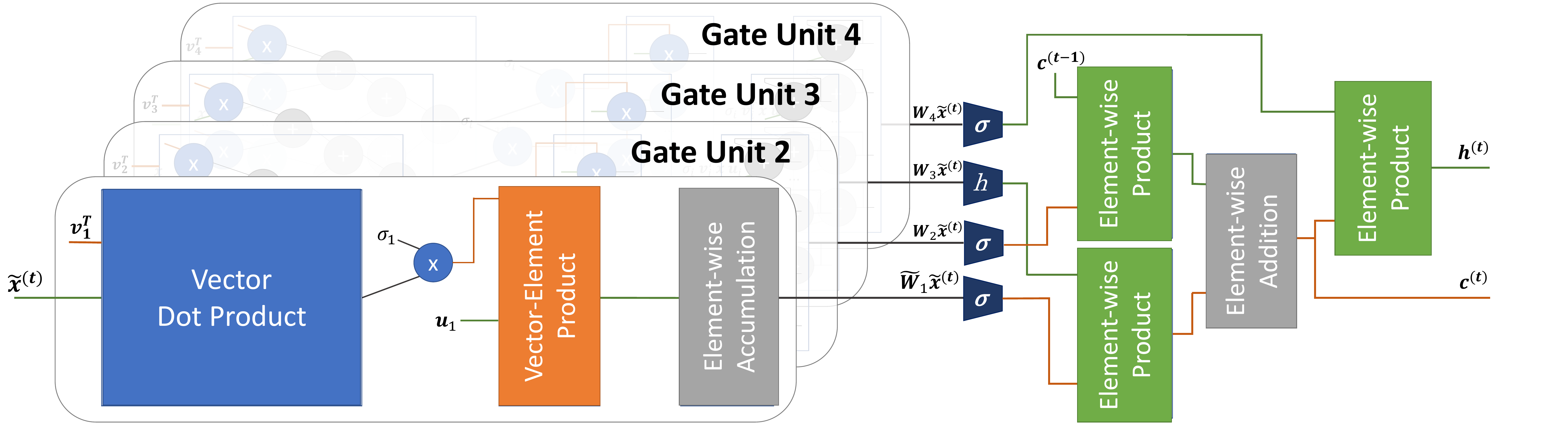}
  \vspace{-0.5cm}
  \caption{ Custom LSTM accelerator architecture (see \cite{rizakis18approximate}).}
  \label{fig:arch}
  \vspace{-0.5cm}
\end{figure}

\textbf{Inter- and intra-gate parallelism.}
Fig. \ref{fig:arch} shows the block diagram of the architecture.  At its core, the architecture is organised as a pipeline of five coarse stages,  including four parallel \textit{hardware gate units}, a set of nonlinear operators and a number of multiplier and adder arrays. Starting on the left-hand side, the four parallel hardware gate units are the heart of the architecture. The proposed design exploits the coarse-grained, inter-gate parallelism by mapping each LSTM gate to a dedicated hardware gate unit, with all units operating concurrently. At each LSTM time-step $t$, a hardware gate unit computes its output by performing $N_{\text{steps}}$ refinement iterations. As a first step, the current input vector is sent from the off-chip memory into an on-chip buffer as it will be reused across all refinement iterations. In the n-th iteration, the singular vectors $\textbf{u}_1^{i(n)}$ and $\textbf{v}_1^{i(n)}$ for the i-th gate are streamed in from the off-chip memory in a tiled manner with tile sizes $T_r$ and $T_c$ respectively, along with the singular \mbox{values $\sigma_1^{i(n)}$}.

Internally, each hardware gate unit contains three processing modules: a\textit{ dot-product} unit, a \textit{multiplier} array and \textit{adder} array (Fig. \ref{fig:arch}). By mapping the operations of a gate to parallel circuits, the architecture capitalises on the fine-grained, intra-gate parallelism of these operations to obtain performance gains. After the hardware gate units have applied all the necessary refinements, the outputs of the four gates are passed through nonlinear operators. Consequently, the produced outputs are processed using the multiplier and adder arrays to produce the new cell state $\textbf{c}^{(t)}$ and output vector $\textbf{h}^{(t)}$.

% Each module can be tunably scaled to provide a performance-resource cost trade-off. 

\textbf{Configurable scaling.}
At compile time, the configuration of the architecture is controlled by means of two parameters: \textbf{\mbox{$T_r \in [1, R]$}} and $T_c \in [1, \text{NZ}]$. $T_r$ controls the size of all the arrays, while $T_c$ determines the number of multiply-add operators in each hardware gate unit. Different values of $T_r$ and $T_c$ correspond to different scaling of the architecture and provide a tunable performance-resource cost trade-off which is used to customise the design based on the available resources and the response-time requirements. 

%TODO - Discuss the selection of the parameter values?

%The main strategy of the FPGA architecture includes the exploitation of the coarse-grained parallelism between the four LSTM gates and is parametrised with respect to the fine-grained parallelism in the dot-product and elementwise operations of the LSTM, allowing for a compile-time tunable performance-resource trade-off.

%At the beginning of the time-step, the current vector $\Tilde{\textbf{x}}^{(t)}$ is stored on-chip as it will be reused in each iteration of all four gates. The vecotrs $\textbf{u}_1^{i(n)}$ and $\textbf{v}_1^{i(n)}$ for each gate, along with the singluar values $\sigma_1^{i(n)}$, are streamed in the accelerator from the off-chip memory in a tiled manner.

\vspace{-2mm}
\section{Navigating the design space}
\vspace{0mm}
\label{sec:dse}

Given an LSTM and a target FPGA, the parameters of the overall methodology comprise the approximation method parameters, NZ and $N_{\text{steps}}$, and the architectural parameters, $T_r$ and $T_c$. Different combinations of these parameters correspond to alternative designs. For a fixed-time constraint, each candidate design is characterised by its: 1) quality of result (QoR), 2) performance in terms of \mbox{processing speed} and 3) resource consumption. To explore this space, we need to study the effect of the architectural parameters on the performance of the hardware implementation as well as the impact of the approximations on the QoR of the target application.

% As first step, we need a metric that captures the impact of the applied approximations on the application-level quality of result for different (NZ, $N_{\text{steps}}$) pairs.
%TODO - In this section, we focus on the problem setup (i.e. NZ, Nsteps, Tc, Tr). We introduce the roofline model in a high level (for example why we need modelling - e.g. avoid long simulations, etc., estimate the performance of a large space of alternative designs, etc.) and state that we developed a roofline model for the proposed architecture. We could focus on how the four aforementioned parameters affect the roofline model, rather than show equations for perf and CTC. We should also mention that we used complete enumeration to find the highest performing configuration. 
\vspace{-3mm}
\subsection{Performance: Following the Roofline}
To investigate the attainable performance of different architectural configurations, we adopt the roofline model \cite{williams2009roofline} from the high-performance computing (HPC) community. The roofline model is a visual model for identifying the causes of performance bottlenecks in computing systems. Based on this model, the performance of a design can be limited by either the peak processing rate of the target platform or by the maximum bandwidth that the external memory subsystem can support. 

% The $y$-axis of the roofline model is the performance in floating-point operations per second, with a flat line indicating the peak processing rate supported by the target platform. The $x$-axis displays the operational intensity, measured as floating-point operations per byte accesses from the external memory.

In this context, we built a roofline model for the proposed architecture which can be used to explore the performance of a large space of alternative designs, without the need for long simulations \cite{rizakis18approximate}. The various candidate designs differ in terms of number of refinement iterations ($N_{\text{steps}}$), level of pruning (NZ) and scaling of the hardware ($T_r$, $T_c$). Given the pruning level NZ, the number of refinements $N_{\text{steps}}$ and a pair of architectural parameters $(T_r, T_c)$, the \textit{attainable performance} of the architecture (in GOp/s) can be modelled as the operation number to latency ratio for each LSTM inference.
\begin{comment}
\begin{equation}
    \label{equ:perf_model}
    \text{Performance} = \frac{\text{ops per LSTM inference(NZ, $N_{\text{steps}}$)}}{\text{latency per sample(NZ, $N_{\text{steps}}$, $T_r$, $T_c$)}}
\end{equation}
\end{comment}

As the weights of an LSTM do not typically fit in the on-chip memory of an FPGA, we model \textit{operational intensity}, also referred to as computation-to-communication ratio (CTC),  as multiplication and addition operations per byte of weights accessed from the external memory (GOp/byte).  % and calculate it as:
\begin{comment}
\begin{equation}
    \label{equ:op_intensity}
    \text{CTC} = \frac{\text{ops per LSTM inference}(\text{NZ}, N_{\text{steps}})}{\text{bytes accessed}(\text{NZ}, N_{\text{steps}})}
\end{equation}
\end{comment}
%where the external memory transfers include the singular vectors and the singular value for each iteration of each gate along with the write-back of the new cell state and the output vector. 
Utilising the above scheme, a design space exploration is conducted to obtain the highest performing set of parameters for both the approximation method and the architecture given the target platform.

% The number of design points enables us to perform a complete enumeration of all candidate designs and in this way tune both the approximation scheme and the architecture with the highest performing set of parameters.

%\begin{comment}
\begin{figure}[t]
%   \vspace{-2mm}
  \centering
  \includegraphics[trim={0.2cm 4cm 0.2cm 1.5cm}
  ,clip,width=0.5\textwidth]{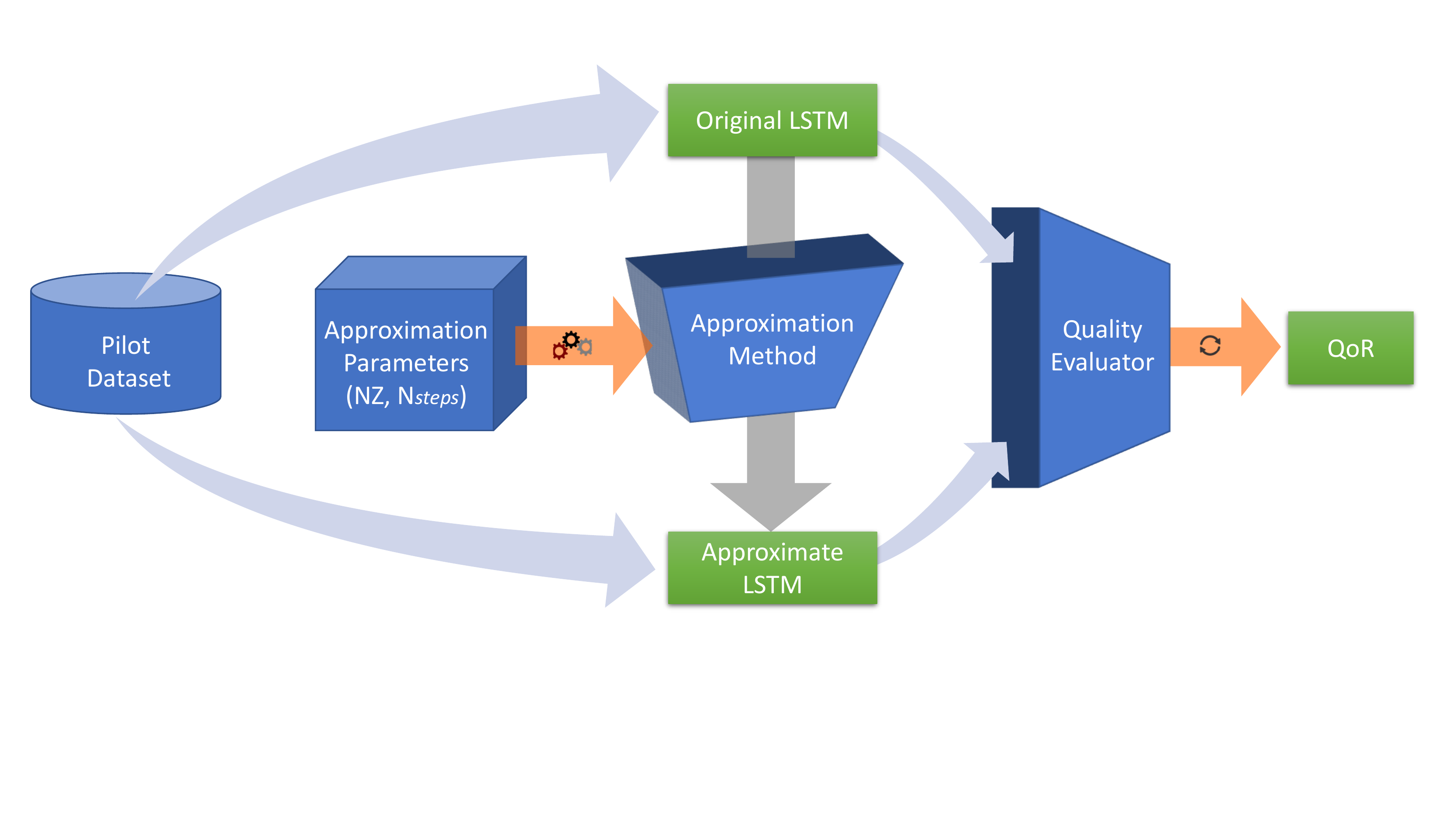}
  \vspace{-0.8cm}
  \caption{Process of capturing the approximation-QoR trade-off.}
  \label{fig:approx_impact}
  \vspace{-0.5cm}
\end{figure}
%\end{comment}

\vspace{-3mm}
\subsection{Level of Approximation vs. Quality of Result}
\label{sec:QoR}
Typically, approximation methods exploit the error tolerance of an application together with the perceptual limitations of humans to trade off quality of result (QoR) with faster processing. Nevertheless, emerging mission-critical systems, such as driverless cars, place safety and robustness at the forefront and hence require guarantees with respect to both QoR and processing latency \cite{McAllister_2017}. To make principled design decisions that meet the requirements of such applications, it is essential to capture the relationship between application-level QoR and level of approximation and use it to tune the computing system based on the application specifications. 

To achieve that, we follow the methodology shown in Fig. \ref{fig:approx_impact}. Initially, the error induced by the proposed LSTM approximations on an application is experimentally measured as a function of the targeted iterations. Given a \mbox{(NZ, $N_{\text{steps}}$)} pair, the approximate LSTM is generated from the original LSTM (top to bottom of Fig. \ref{fig:approx_impact}). Next, we run the target application end-to-end over a pilot dataset using both the original and the approximate LSTM. By treating the final output of the original model as the ground truth, an application-specific metric is employed to assess the QoR of the approximate LSTM (left to right of Fig. \ref{fig:approx_impact}). The quality metric measures the similarity between the original and the approximate result and must have a suitable form based on the target domain, such as the relative error between the approximate and reference result or the Kullback-Leibler (KL) divergence that captures the distance between the respective probability distributions. Overall, by varying the values of (NZ, $N_{\text{steps}}$) and observing the associated QoR, the relationship between the level of approximation and the QoR is captured.

\vspace{-2.5mm}
\section{Case study: autonomous driving}
\subsection{Overview}
One of the emerging AI-driven applications with the highest potential for societal impact is autonomous driving. Although initial efforts begun in the late 1980s \cite{pomerleau1989alvinn}, the field of autonomous driving has experienced significant progress in the past decade, owing to efforts from both the industrial and academic communities. The main enablers of the emerging technologies being developed are: (i) the advancement of deep learning algorithms allowing the extraction of powerful representations, (ii) the availability of real-world training data provided by open-source datasets \cite{geiger2012we,yu2018bdd100k} and (iii) the development of embedded processing platforms with enhanced computational capabilities that allow the deployment of computationally expensive software on-board the vehicle \cite{liu2017computer} \cite{ce3}, satisfying the imposed low-latency and safety constraints. 

Vision-based driving assistance and autonomy \cite{bojarski2016end,chen2015deepdriving} \cite{ce7} is gaining attention due to the low-cost, widely available cameras that can be used independently or accompany other sensors %such as LiDAR and Sonar 
for environmental perception. With such sensors providing a stream of measurements, recurrent models such as LSTMs form a promising learning paradigm that can extract and exploit temporal information from the incoming data to develop a smooth and consistent driving policy, in place of the independent per-input predictions provided by classical deep learning models that exploit solely spatial information \cite{chi2017deep}. %In Fig. \ref{fig:autonDriving} for example, a human is not able to infer any insights regarding the front vehicle's velocity or trajectory by visually observing the given image. Contrary, by observing a sequence of such frames it is possible to reason about the vehicles current state and predict its future behaviour.

Self-driving car systems consist of a large set of computationally demanding tasks, including sensor preprocessing, localisation, mapping, path planning and obstacle avoidance, control and emergency handling \cite{thrun2010toward}. Hard low-latency constraints \cite{ce1} between perception and action impose the need for high-performance implementations that guarantee the extraction of highly accurate approximations on each individual component, to meet the real-time performance requirements of the overall system with insignificant effect on accuracy. As an example, a coarse but in-time estimation of the vehicle's obstacle avoidance system to take a ``sharp'' left turn and avoid a collision, is preferred to a delayed but rather accurate regression of an exact steering angle response to a visual input.  

\textbf{Target Application.} The driving model presented in \cite{Xu_2017}, trained on the Berkeley DeepDrive Video dataset (BDDV), a large-scale crowdsourced driving video dataset forming an early version of the BDD100K Dataset \cite{yu2018bdd100k}, is examined as a case study for evaluating the proposed framework. Similar to the work of \cite{ce5} on vision-based autonomous mobile robot navigation, Xu et al. also exploit the end-to-end learning paradigm.  Input frames for each video are first processed by a Fully-Convolutional Network (FCN) to encode the spatial features which are then fed to a trained LSTM model that predicts the probability distribution across a discrete set of feasible future actions for the vehicle (go forward, stop, turn left, turn right) taking advantage of the temporal motion information from previous representations. The LSTM input is enhanced with the linear and angular velocities of the vehicle predicted by the system from the previous frame. This FCN-LSTM architecture is a novel version of Long-term Recurrent Convolutional Networks (LRCNs), typically consisting of a convolutional neural network feeding its output to an LSTM, combining the current state-of-the-art in visual and sequence learning to extract spatio-temporal information for input streams. %By replacing the CNN with a FCN, the learned visual representation preserves the locality of information, while the end-to-end network is trained  jointly with a side task of semantic segmentation undertaken by the FCN. 
\subsection{Evaluation}
\vspace{-0mm}
 In this section we discuss the extensive experimental evaluation conducted to showcase the effectiveness of the proposed approach in the target application of this case study. The proposed progressive inference methodology is initially compared with an FPGA-based baseline for LSTM inference to demonstrate its efficacy on making informed predictions under computation time constraints (Sec. \ref{sec:fpga-comparison}). Then, a comparison of the proposed methodology %implemented on the introduced FPGA-based custom hardware architecture,
with faithful off-the-self LSTM implementations targeting other computing platforms (CPU and GPU) considering latency, power consumption and performance efficiency is discussed \mbox{(Sec. \ref{sec:gpu_comparison}).} 

\textbf{Experimental Setup.} We focus on the LSTM of the examined driving model for this case study, each gate of which %consists of two $R \times C$ weight matrices,
forms an $R \times C$ augmented weight matrix, with $R=64$ and \mbox{$C=8320$}. We evaluate the method on part of the validation set of the dataset that was used to train the model, by cropping a segment of 100 consequent frames from over 1800 real videos of diverse driving scenarios. % Ground truth labels are obtained using GPS/IMU data from the vehicle.    
To generate action probability distributions that will act as ground truth for the evaluation of the proposed approximation method, we follow the process of Sec. \ref{sec:QoR} and execute the original driving model end-to-end over the validation set using TensorFlow. As a metric of the effect of low-rank approximation and prunning on the QoR, we employ Kullback-Leibler (KL) divergence -a commonly used metric of dissimilarity between distributions- between the reference and predicted probability distribution.

In our experiments, we target the Xilinx's ZC706 board mounting the Zynq 7045 chip. This platform is an industry standard for FPGA-based embedded systems and is based on the Zynq-7000 System-on-Chip which integrates a dual-core Arm CPU alongside an FPGA fabric on the same chip. For the data format, we use single-precision floating-point representation to comply with the typical precision requirements of LSTMs as used by the deep learning community. All hardware designs are synthesised with Vivado HLS and Vivado Design Suite (v2017.1) achieving a clock frequency of 100 MHz.

\begin{figure}
%   \vspace{-2mm}
  \centering
  \includegraphics[trim={20mm 38mm 20mm 46mm },clip,width=0.5\textwidth]{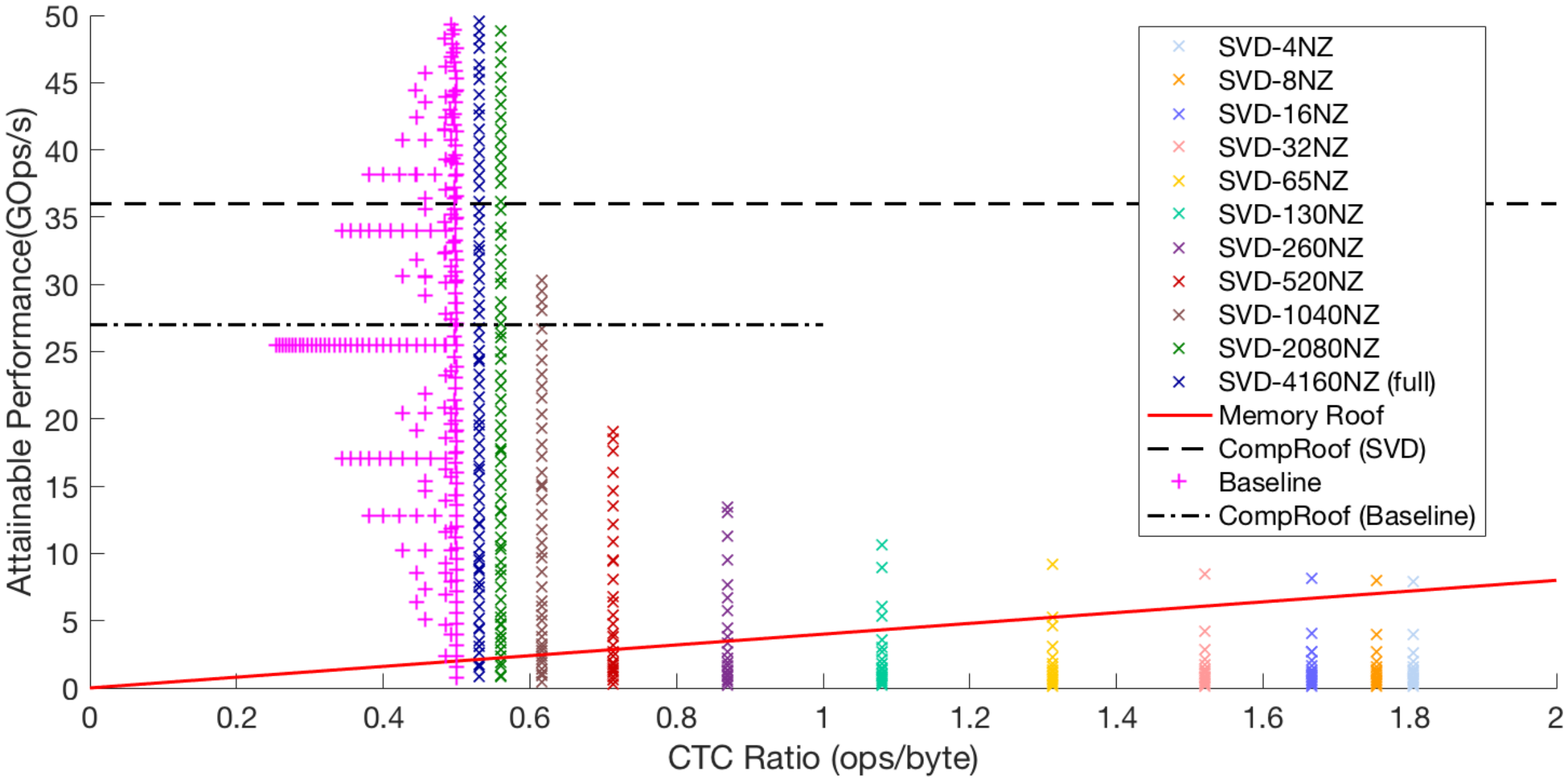}
  \put (-228,20) {\footnotesize \color{red}4 GB/s}
  %\vspace{-0.5cm}
  \caption{Roofline model analysis for the baseline architecture and various configurations of the proposed method.}
  \label{fig:roofline}
  \vspace{-0.5cm}
\end{figure}

The core LSTM workload of the proposed approximate computing scheme (dot-product followed by a vector scaling by a constant), as well as the baseline LSTM implementation of Sec. \ref{sec:fpga-comparison} (matrix-vector multiplication), is implemented on the FPGA.  At the same time, the CPU coordinates the operation of the system by: (i) scheduling the computations between different tiles from all 4 LSTM gates and mapping them to the available processing elements of the custom hardware accelerator, and (ii) setting up the communication interface between the accelerator and the external memory. To this end, we use the four AXI-based High-Performance (HP) ports that are available on the target device. For each port, we configure it with a 64-bit width and instantiate a dedicated DMA engine, clocked at 150 MHz, to independently perform the memory transfers. Overall, our memory interface subsystem yields a measured bandwidth of around 4 GB/s as shown on the slope of the roofline model in Fig. \ref{fig:roofline}, with the CPU initialising the DMA engines state prior to execution.

%memory transfers through the AXI interface to efficiently utilise the available memory bandwidth.

%Stelios: TODO-me: Check HLS code by Michalis. I think the memory transfers are done in bursts via a DMA engine (need to check this). If this holds, then the CPU sets up the DMA engine, initialises the accelerator and launches the execution. We can add details on the memory ports that we use, i.e. (4? 64-bit) High-Performance (HP) ports of the Zynq SoC, clocked at 150 MHz - approaching a bandwidth of 4 x 8 x 150M 1e-6 GB/s ~ 4 GB/s (also shown on the slope of the roofline model (ARC 2018)).

In the comparison of the proposed methodology with a CPU- and GPU-based LSTM implementation (Sec. \ref{sec:gpu_comparison}), we used PyTorch (v1.1.0) with CUDA 10, to develop a faithful LSTM baseline and deploy it on the widely used NVIDIA Jetson AGX Xavier board (which was also presented at the 2017 Consumer Electronics Show \cite{ce8}), featuring an 8-core Arm 64-bit CPU along with a 512-core Volta GPU. Average performance and power are calculated after completing 1000 iterations of each experiment across all platforms. The idle power is subtracted from all measurements, leading to a comparison of the actual power consumed by the benchmark execution (including the memory accesses). 

%TODO in Future Version: Incorporate 4GB/s Memory roof in the matlab plot. (Currently superimposed in latex figure as text.) 

%\textbf{Comparison with FPGA baseline.}
\subsubsection{\textbf{Comparison with FPGA baseline}}
\label{sec:fpga-comparison}
\begin{figure}
  \vspace{-4mm}
  \centering
 \begin{tabular}{@{}c@{}}
  \footnotesize{(a)}\includegraphics[trim={60mm 35mm 60mm 38mm },clip,width=0.45\textwidth]{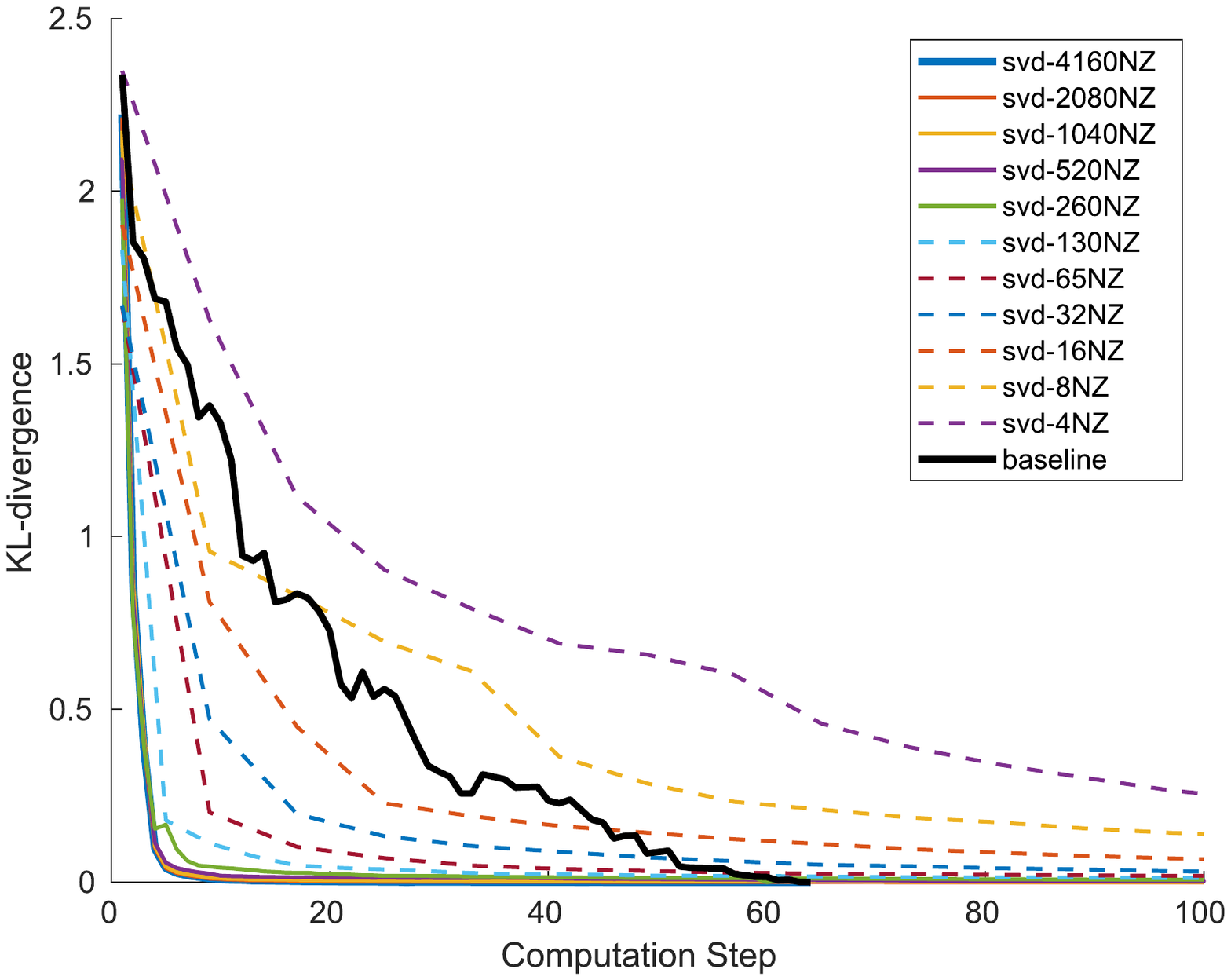} \\
  %\begin{flushleft} \small (a) An image \end{flushleft}
  \vspace{-0.62cm}
  %\caption{KL-divergence between  approximate prediction at each computation step and reference model output.}
\end{tabular}
\begin{tabular}{@{}c@{}}
%   \vspace{-2mm}
  %\centering
    \footnotesize{(b)}\includegraphics[trim={60mm 32mm 60mm 35mm },clip,width=0.43\textwidth]{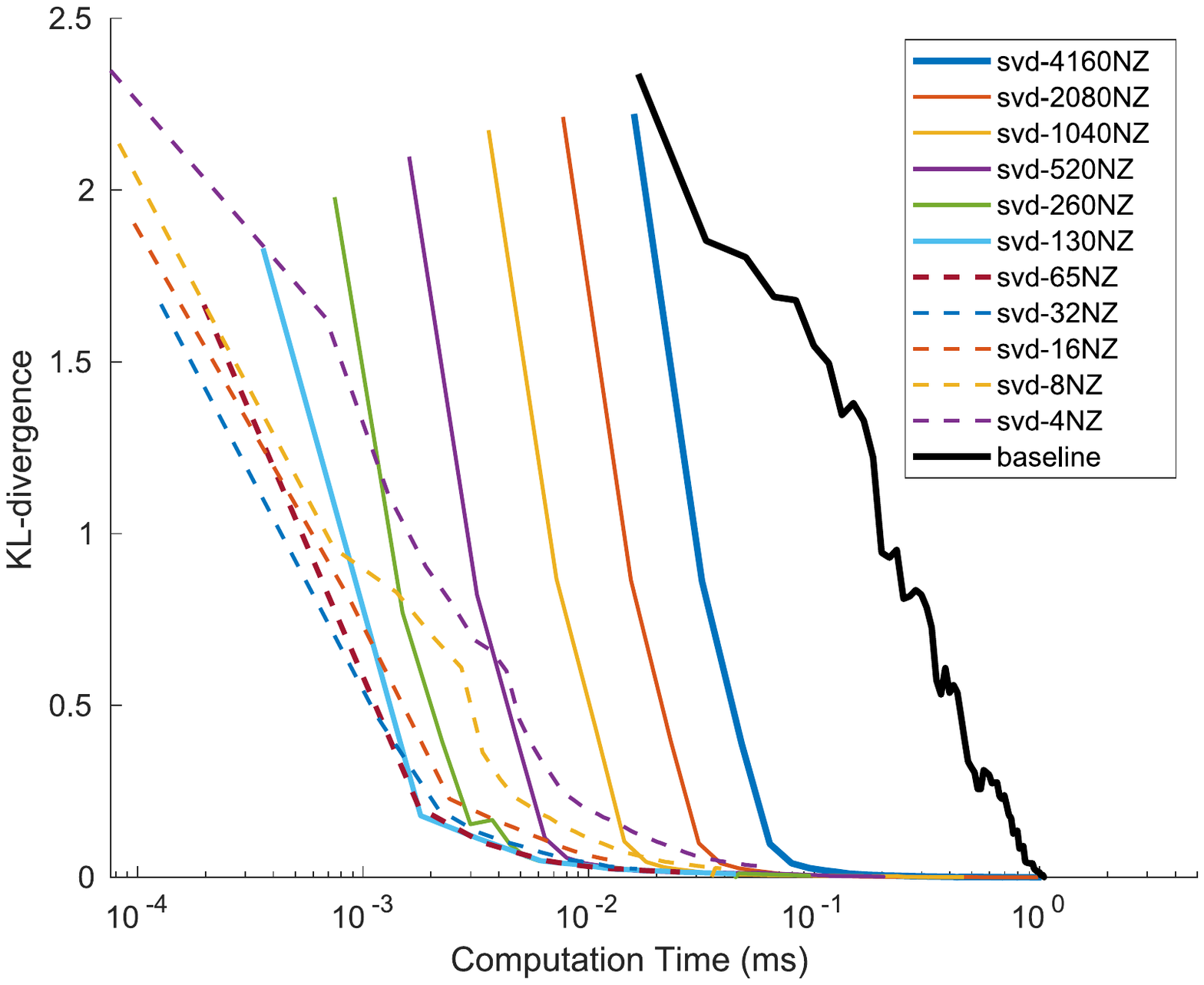}
  %\vspace{-0.4cm}
  %\caption{KL-divergence between approximate prediction and reference model output as a function of computation time.}
  \vspace{-0.2cm}
\end{tabular}
  \caption{KL-divergence between approximate prediction and reference model output (lower-is-better) as a function of: (a) Computation Step, (b) Computation Time.}
  \label{fig:comparison}
    %\vspace{-0.4cm}
\end{figure}

A hardware architecture implementing a faithful mapping of the original LSTM model described in Sec. \ref{sec:background} is developed to act as a baseline for the evaluation of the proposed system. This baseline architecture consists of four gate units with a total of 2.1M parameters, implemented in parallel hardware that performs the matrix-vector multiplication operations of LSTM gates (Fig. \ref{fig:lstm}) in a blocked manner. The computational workload for the kernel of each gate is $2RC$ operations. Parametrisation with respect to the tiling along the rows ($T_r)$ and columns ($T_c$) of the weight matrices is applied and roofline modelling is used to obtain the highest performing configuration \mbox{($T_r$, $T_c$)}, similarly to the proposed system's architecture (Fig. \ref{fig:roofline}). %The maximum platform-supported attainable performance was obtained for $T_r=1$ and $T_c = 4$. %, utilising X DSPs (X\%), X kLUTs (X\%), X kFFs (X\%) and \mbox{X 18kbit BRAMs (X\%)}. 
As \mbox{Fig. \ref{fig:roofline}} demonstrates, the designs are mainly memory-bound and as a result a small portion of the FPGA resources are utilised. To obtain the application-level QoR of the baseline design under time-constrained scenarios, the KL divergence between the intermediate LSTM output at each tile step of $T_r$ and the predictions of the reference model is examined and illustrated by the black line of Fig. \ref{fig:comparison}b.

\begin{figure*}
  \vspace{-5mm}
  \centering
   \includegraphics[trim={38mm 40mm 30mm 37mm },clip,width=0.85\textwidth]{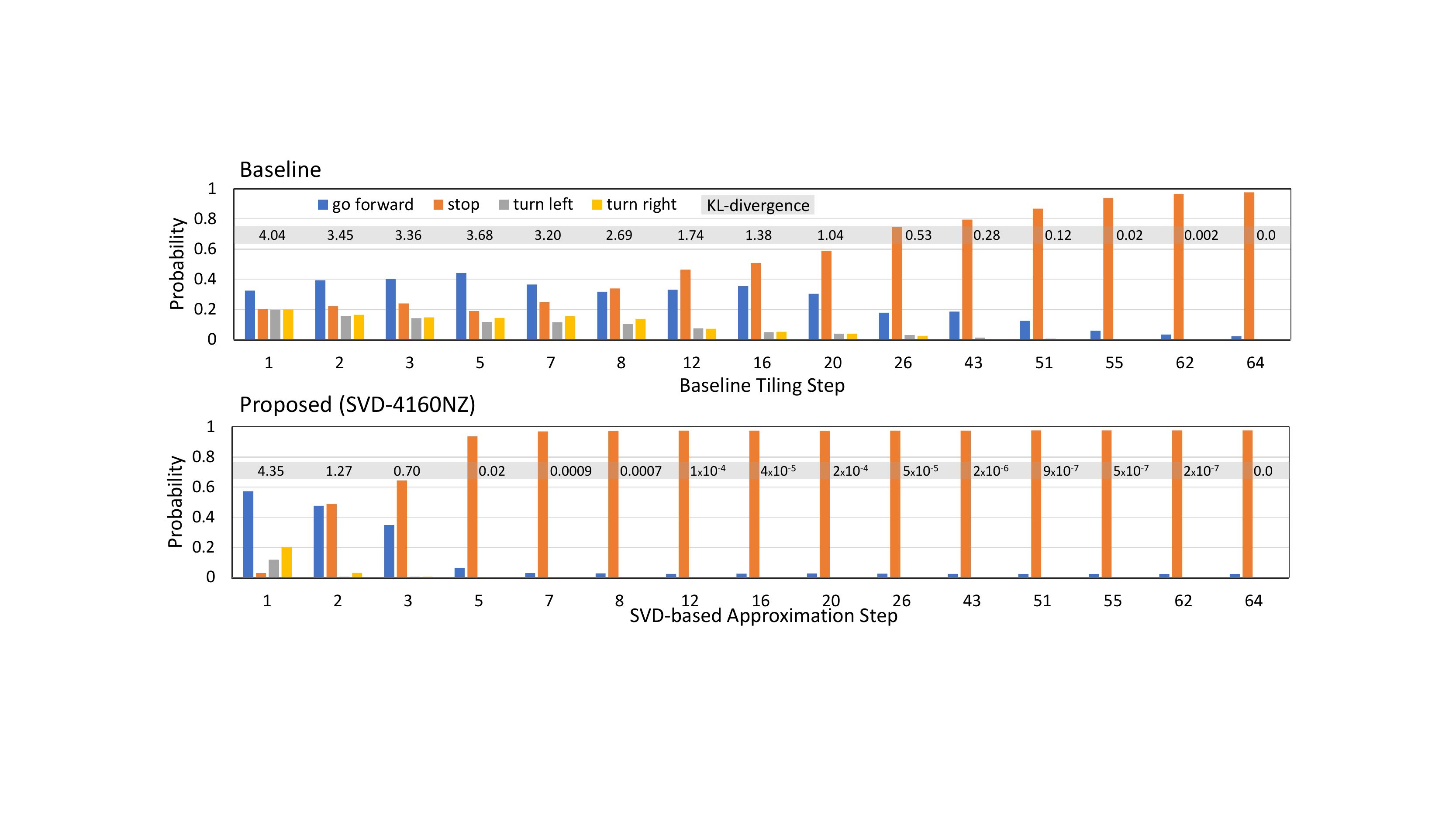}
         \vspace{-2mm}
   \caption{ Intermediate prediction instances obtained by the progressive inference baseline (achieving $16.6^{-2}$ms/step) and a dense instance of the proposed SVD-based approach (achieving $15.9^{-2}$ms/step) on the same data sample, as a function of computation steps. KL-divergence values with respect to the final result are also shown (grey row).}
    \label{fig:timelapse}
      \vspace{-5mm}
\end{figure*}

The gains of the proposed methodology compared to the baseline design under computation-time constraints are investigated by exploring the design space, defined by (NZ, $T_r$, $T_c$), in terms of \mbox{(i) performance} (Fig. \ref{fig:roofline}) and (ii) the relationship between error (described by the KL-divergence between the approximate prediction and ground truth) and computation time \mbox{(Fig. \ref{fig:comparison}b).} \mbox{Fig. \ref{fig:comparison}a} also depicts the relationship between error and computation step for numerous configurations of the proposed system. As illustrated, the QoR of a configuration is inversely proportional to its level of sparsity. Dense configurations, such as those with 50\% non-zero elements or more, tend to converge to negligible divergence values \mbox{(below $10^{-6}$)} in less than 15 computation steps, in contrast with sparser configurations that require more than 75 computations steps to converge to the same divergence level (\texttildelow 2\% non-zero elements) or converge to higher divergence values (as in the case of 0.4\% non-zero elements). Additionally, Fig. \ref{fig:timelapse} presents probability distribution instance samples of numerous progressive refinement steps for a representative input frame along with their corresponding KL-divergence values. It can be seen that the proposed approach convergences to ``meaningful results" (application-wise) in much smaller number of computation steps, by exploiting the inherent redundancy of the LSTM model. % highlighting the improved QoR as a function of time budget.

% sample instances of intermediate probability distributions corresponding to progressive approximation iterations for a representative input frame, are illustrated in Fig. \ref{fig:sampleSteps}, showing the improved QoR as a function of time budget.

\begin{figure}
  \centering
  \includegraphics[trim={0mm 0mm 0mm 0mm },clip,width=0.5\textwidth]{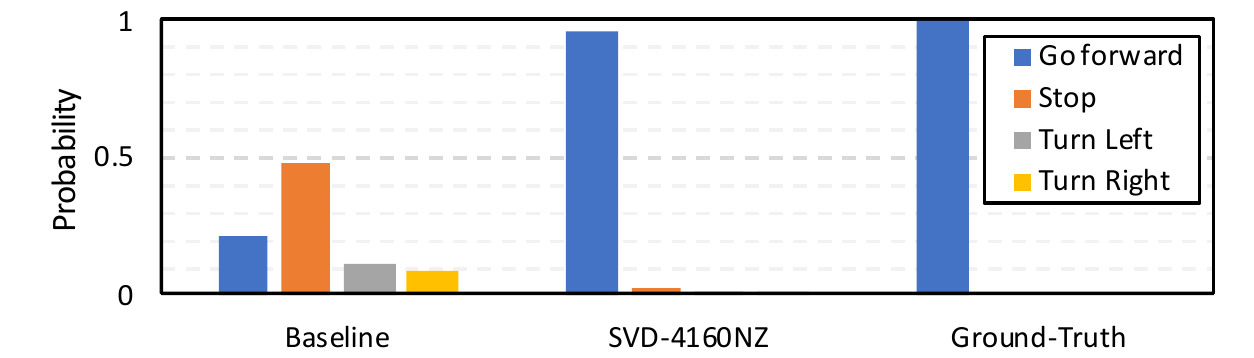}
   \caption{Intermediate prediction instances obtained by the baseline and the proposed approach with NZ=4160 on the same data sample, under the same latency constraint (t=$10^{-1}$ms).}
    \label{fig:instances}
      \vspace{-5mm}
\end{figure}

As shown in Fig. \ref{fig:comparison}b, since computation time per computation step is also inversely proportional to the level of sparsity of a given configuration, some sparse configurations demonstrate superior accuracy than other denser settings under the same latency constraint. This behaviour, however, is not monotonic due to extremely dense configurations requiring a larger number of computation steps to converge. Therefore, the selection of the appropriate level of sparsity is dependent on the latency constraint imposed by the application-level needs.  Overall, we notice that the proposed methodology achieves a speed-up of 198$\times$ on average (76$\times$ geo. mean) across different quality-of-result levels compared to the baseline approach. 
In particular, when only negligible KL-divergence is allowed between the approximate and reference prediction, the proposed system achieves 2.93$\times$ faster inference by exploiting the LSTM model's inherent redundancy. Furthermore, the proposed method demonstrates up to 415$\times$ lower inference time to achieve an intermediate QoR prediction exploiting the computation time-accuracy trade-off.
%Stelios: Thanks for adding this, the discussion is good - these are the points we want make. I just think we can improve a bit the clarity, because there's a lot to digest. I rephrased a bit above, but will come back to it.
%This improvement ranges from 2.63$\times$ faster inference, arising directly by exploiting the reference LSTM model's redundancy in the case where only negligible KL-divergence is allowed between the approximate and the reference prediction, up to 237$\times$ lower inference time required to achieve an intermediate QoR prediction exploiting the computation time-accuracy trade-off.
%
 Fig. \ref{fig:instances} illustrates two representative intermediate probability distributions extracted by an instance of the proposed approach and the baseline, both satisfying the same latency constraint. To obtain these outputs, both methods were fed with the same input and while calculating their predictions their computation was cut short as the available time budget was hit. The illustrated intermediate output distributions indicate that the proposed approach makes a more informed prediction, significantly closer to the ground-truth compared to the baseline. This property is particularly useful in scenarios where tight real-time requirements impose hard latency constraints on the available computation time budget for inference.

\begin{table}
\caption{Comparison with other computing platforms}
\vspace{-3mm}
\begin{center}
\renewcommand{\arraystretch}{1.2} 
\setlength\tabcolsep{4pt} % default value: 6pt
{ \color{black}
\resizebox{1\columnwidth}{!}{

\begin{tabular}{ ccccc }
 \hline
 \textbf{ Platform} & \textbf{Benchmark} & \textbf{Latency} & \textbf{Power} & \textbf{Perf. Efficiency} \\
 \hline
 CPU & Baseline  & 2.4266 ms & 5.13 W & \phantom{00}0.342 GOp/s/W \\ 
 GPU & Baseline  & 0.7974 ms & 7.11 W & \phantom{00}0.752 GOp/s/W \\ 
 FPGA & Baseline  & 1.0620 ms & 2.72 W & \phantom{00}1.476 GOp/s/W \\ 
 \hline
 FPGA &  $\dagger^{*}$ (KL$\leq0.001$) &0.36190 ms & 2.76 W  &  \phantom{000}4.267 GOp/s/W \\ 
 FPGA &  $\dagger^{**}$              (KL$\leq0.01$)& 0.03924 ms & 3.20 W  &  \phantom{00}33.943 GOp/s/W \\ 
 FPGA &  $\dagger^{**}$\phantom{*}              (KL$\leq0.1$) & 0.00335 ms & 3.20 W  &  \phantom{0}397.713 GOp/s/W \\ 
 FPGA &  $\dagger^{***}$\phantom{}                (KL$\leq1.0$) & 0.00072 ms & 3.41 W  & 1735.992 GOp/s/W \\ 
 \hline
   \multicolumn{5}{c}{
%       \begin{flushleft} 
         $\dagger$ This work, $^{*}$SVD-4160NZ (no pruning), $^{**}$SVD-130NZ, $^{***}$SVD-32NZ 
%       \end{flushleft}
   }
\end{tabular}
}
}
\vspace{-0.6cm}
\end{center}
\label{tab:other_platforms} 
\end{table}

\begin{comment}
\begin{figure}
  \vspace{-5mm}
  \centering
   \begin{tabular}{@{}c@{}}
  \footnotesize{(a)} \includegraphics[trim={0mm 0mm 0mm 0mm },clip,width=0.5\textwidth]{sample_v6.pdf}
    % \caption{Visualisation of intermediate prediction instances obtained by the proposed approach with 8 NZ elements.}
  %\label{fig:sampleSteps}
\end{tabular}
\begin{tabular}{@{}c@{}}
   \footnotesize{(b)} \includegraphics[trim={0mm 0mm 0mm 0mm },clip,width=0.5\textwidth]{base_svd_gt_v4.pdf}
  %\caption{Comparison between predictions of the baseline and the proposed approach with 4160 NZ elements, under the same latency constraint.}
  %\label{fig:base_svd}
\end{tabular}
   \caption{Intermediate prediction instances obtained by: (a) the proposed approximation scheme with NZ=8 in various time-steps of the computation, (b) the baseline and the proposed approach with NZ=4160, under the same latency constraint (t=$10^{-1}$ms).}
    \label{fig:instances}
      \vspace{-5mm}
\end{figure}
\end{comment}

\subsubsection{\textbf{Comparison with CPU and GPU baselines}}
\label{sec:gpu_comparison}
Targeting the efficient deployment on the embedded space, deep-learning models should abide in a low-power envelope. Power efficiency becomes increasingly prominent in the case of autonomous systems \cite{ce7} that rely on self-contained power supply resources, and especially in self-driving cars that are emerging alongside with the rise of the electric vehicle era. Power-constrained applications are primarily concerned about: (i) the absolute power consumption (watts) and (ii) the performance efficiency (performance-per-watt). 

In this respect, we also compare multiple instances of the proposed methodology and its underlying FPGA-based hardware implementation with highly optimised off-the-shelf CPU- and GPU-based traditional implementations of LSTM inference, commonly used by the deep learning community, in terms of raw performance, absolute power consumption and performance efficiency. Although raw performance and power consumption are also reported, the most equitable metric for cross-platform comparison is power efficiency, as it effectively normalises the results with respect to the available computational resources of each target platform. 

Table \ref{tab:other_platforms} summarises the results of this comparison. The employed FPGA baseline achieves a 2.28$\times$ speed-up compared to the CPU implementation, while suffering a 0.75$\times$ slow-down with respect to the GPU, in terms of absolute latency. However, when power consumption is also considered, these results are translated into a 4.31$\times$ and 1.96$\times$ improvement on performance efficiency compared to the CPU and GPU baseline respectively. These demonstrated gains in power efficiency achieved by the use of a custom FPGA-based solution render FPGAs as the cardinal platform for LSTM deployment in many power-constrained applications, especially in the embedded space of autonomous systems. 

Multiple instances of the proposed approximate computing scheme are also listed in Table \ref{tab:other_platforms}. It can be seen that by utilising solely the proposed computation-restructuring methodology, a speed-up of 6.7$\times$, 2.2$\times$ and 2.93$\times$ is achieved in the latency required to yield (almost) identical outputs \mbox{(KL-divergence$\leq0.001$)} with the reference design, compared to the CPU, GPU and FPGA baselines accordingly, also translated into an improvement of 12.46$\times$, 5.67$\times$ and 2.89$\times$ in performance efficiency. These significant gains arise by the proposed methodology exploiting the inherent redundancy of LSTM models in order to maximise the achievable accuracy at every stage of the computation. By performing the most information-carrying computations first, the workload (and computation time) required to reach similar accuracy with the baseline is effectively reduced. 

By relaxing the error tolerance into slightly higher KL-divergence values ($\leq0.1$), the proposed hybrid compression-and-pruning methodology provides informed approximations of the inference outputs, while demonstrating remarkable performance gains of up to 724$\times$, 238$\times$ and 317$\times$ in latency (1161$\times$, 529$\times$ and 269$\times$ improved performance efficiency) compared to the same CPU, GPU and FPGA baselines respectively. These gains are amplified remarkably by further relaxing the error tolerance into higher KL-divergence ($\leq1.0$), which however still yield ``meaningful" results.

Since all the configurations of the proposed (and baseline) approach are memory bounded (Fig. \ref{fig:roofline}), the attainable parallelism and performance (GOp/s) of the underlying hardware architecture increase proportionally to the selected sparsity level. As it can be noticed in Table \ref{tab:other_platforms}, the absolute power consumption of sparser configurations increases. This is due to the fact that since sparser configurations lead to higher CTC ratio (Fig. \ref{fig:roofline}), more parallel processing can be exploited in this case, by instantiating more on-chip computational resources on the FPGA. Consequently, although processing becomes faster, the absolute power of the accelerator also increases as a result of the on-chip power consumption.

% that as a result of this higher exploitation of computational resources due to the increased parallelism, the absolute power consumption of sparser configurations is also increased.

Exploiting the computation time-accuracy trade-off, the proposed progressive inference methodology can provide high-quality approximations of the final result at early stages of the computation, which are iteratively refined as more time budget becomes available. This scheme is particularly useful for systems with hard computation-time constraints (\textit{e.g.} in mission-critical real-time applications), enabling them to maximise the attainable quality of result within the given latency envelope. Furthermore, the introduced highly-parametrised custom hardware architecture for the proposed methodology demonstrates remarkable power efficiency by exploiting the enhanced customisation capabilities and flexibility of FPGAs. In this manner, highly-optimised hardware mappings of different configuration instances of the proposed approximation scheme are generated, while being tailored to the needs of the target application.

\section{Related Work}
\label{sec:related_work}

% 1) FPGA-enabled works, i.e. custom hardware in order to obtain performance gains
%   - ESE (pruned, sparse network) [FPGA 2017], E-RNN (hw/sw co-design) [HPCA 2019], Wang et al. [TVLSI 2017].
%       + Common denominator -> quantisation in non-standard precisions (12-bit and less)
%   - DeltaRNN [FPGA 2018]. However, a similar inference-time-only technique has also been applied for embedded GPUs [MICRO 2018].
%   - Circulant and block-circulant matrices: Complementary to our method, since SVD is still applicable (w/ varyinf perf. gains).
%
% 2) Efficient processing of CNNs.
% 3) Progressive inference
%   - Early-exit CNN classifiers.

The rapid advances in deep learning have lead to significant research effort invested in optimising the execution of deep neural networks. The majority of existing work has focused on compute-intensive convolutional neural networks (CNNs) for computer vision tasks. The substantial redundancy of modern deep CNNs together with the inherent parallelism and data-reuse of CNN workloads have made them amenable to various compression and acceleration techniques. At the algorithmic level, methods such as knowledge distillation \cite{distill_2015}, efficient convolutions \cite{cheap_conv_2018} and neural architecture search \cite{automl_eccv_2018} have been successfully applied to significantly compress CNN models by leveraging their high inherent redundancy. At the same time, techniques such as reduced precision \cite{kouris18cascade, hw_aware_quant_2019_CVPR} and custom hardware designs \cite{sv2018csur} have been employed for acceleration by exploiting the high levels of parallelism and data reuse of CNNs. Nevertheless, with memory-bound LSTMs having substantially different computational patterns, the CNN-centric methods and accelerator designs either provide minimal gains or are not directly applicable to LSTMs \cite{micro_2018, brainwave2018isca}.

% \textbf{Progressive inference via early exits.}
%Progressive inference refers to networks that refine their accuracy as a function of computation time. In this context, closer to the philosophy of our approach lie CNNs that employ early-exit classifiers. 
Closer to the progressive inference philosophy of our approach lie CNNs that employ early-exit classifiers.
CNNs with early exits \cite{branchynet_2016,msdnet_2018,overthink_2019} provide a run-time accuracy-latency trade-off and are able to produce an increasingly refined output as a function of time, which casts them suitable for time-constrained inference scenarios. However, as the early-exit classifiers have to be trained, access to the training set is necessary and complex hyperparameter tuning is required \cite{msdnet_2018,overthink_2019}. Furthermore, although early exiting has been applied to CNN-based classifiers with promising results, this mechanism is not directly applicable to the substantially different topology of LSTMs. Alternatively, our method enables us to perform progressive inference using LSTM models without the need to access the training set and the excessive time overhead of tuning the associated hyperparameters.

%Alex: Should we add Progressive Inference subsection here, as a "most closely related direction to our work" which is also not applicable on LSTMs, before mentioning FPGAs?

With a focus on LSTM workloads, several works have proposed optimisations for executing LSTMs on conventional programmable platforms such as CPUs \cite{deepcpu_atc_2018} and GPUs \cite{persistent_rnns2016icml,sparse_pers_rnns_iclr_2018,micro_2018}. By employing tailor-made caching and data-locality strategies, this line of work has demonstrated significant performance gains and has approached the performance limits of commodity programmable hardware architectures. To push further the attainable performance of LSTMs, another line of work has exploited the characteristics of FPGAs to propose custom accelerator designs. 
%
%Alex: The first sentence here repeats the prev paragraph
% While significant research effort has been invested in optimising the execution of LSTM workloads on conventional programmable platforms such as CPUs \cite{deepcpu_atc_2018} and GPUs \cite{persistent_rnns2016icml,sparse_pers_rnns_iclr_2018,micro_2018}, several works have exploited the characteristics of FPGAs to push further the attainable performance of LSTMs. 
Based on the stage where optimisations are applied, FPGA-based LSTM designs can be categorised into: 1) post-training with fine-tuning, 2) training-stage and 3) run-time methods.

%Alex: I like this categorisation, it may be nice if in each section we first present the algorithmic optimisation and then connect them to what is enabled by FPGAs. (not sure if it is applicable everywhere, but it may be relevant when we have both model-level optimisation and efficient mapping on accelerator)

% . To push farther the attainable performance of LSTMs, several works have exploited the characteristics of FPGAs to design high-performance custom hardware solutions. Based on the stage of the deep learning pipeline that optimisations are applied, FPGA-based LSTM designs can be categorised into: 1) post-training with fine-tuning, 2) training-stage and 3) run-time methods.

\subsubsection{Post-training methods with fine-tuning}
By putting emphasis on minimising the effect of memory boundedness of LSTM workloads, ESE \cite{Han_2017} proposes to sparsify LSTMs via a pruning scheme and map it on an FPGA-based accelerator tailored for sparse workloads. Given a pre-trained model, its weights are pruned in an iterative manner using a load-balance-aware strategy that aims to sustain the utilisation of the accelerator high. Furthermore, to avoid excessive accuracy drop, at each iteration the unpruned weights are fine-tuned using the training set. To overcome the inefficiencies of CPUs and GPUs when executing the sparse, pruned model, ESE exploits the customisability of FPGAs to propose an accelerator optimised for sparse computations. As a result, the load-balance-aware pruning leads to 6.2$\times$ faster execution over dense LSTMs on ESE's accelerator. To further improve the load balancing, Park \textit{et al.} \cite{Park_2019} proposed an alternative encoding format for storing sparse matrices and managed to achieve higher sustained utilisation of the PEs on the same accelerator.

Overall, despite the fact that the pruning method used by both ESE and Part \textit{et al.} is applied \textit{post-training} on pre-trained LSTMs, access to the training set is required in order to iteratively prune and fine-tune the model's weights and thus not significantly degrade the accuracy. In contrast, our method is also applied post-training on pre-trained models, but it does not require access to the dataset and hence is suitable for privacy-aware cases.

\subsubsection{Training-stage methods}
By modifying the model design process, 
Wang \textit{et al.} \cite{Wang_2017} proposed a compression technique that modifies the LSTM model \textit{before} the training stage. By applying a circulant structure to the matrices within each LSTM gate, this approach allows the same weights to be shared across several neurons and substantially reduces model size and storage requirements. Further parametrising this technique, the E-RNN \cite{ernn_2019} framework introduces a blocking version of circulant matrices and treats the block size as a tunable parameter to balance processing speed and accuracy. The block-circulant matrix operations were executed in the frequency domain to leverage the computational efficiency of FFT. At the hardware level, to bypass the limitations of conventional platforms when executing irregular computations, E-RNN proposed a highly-parametrised custom hardware architecture mapped on the flexible FPGA fabrics, leading to a 7.7$\times$ speed-up over \mbox{ESE \cite{Han_2017}}.

In contrast with our post-training approach, both of these methods are applied at the LSTM model level and intervene substantially with the LSTM model design and training. Nevertheless, since the SVD-based decomposition of our work is applicable to circulant matrices, our scheme is orthogonal to these works and can be applied in a complementary manner to yield further performance improvements.
%%%%

% Both of these methods are applied at the model design stage and intervene substantially with the LSTM model design and training. In contrast to our method that is applied post-training without need to access the training set.

% applies a circulant structure to the matrices within LSTM gates. This approach allows the same weights to be shared across several neurons and effectively significantly reduces model size and storage requirements. The E-RNN \cite{ernn_2019} framework further parametrises this technique by introducing a blocking version of circulant matrices with the block size as a tunable parameter. In this manner, the size of a circulant block can be selected to co-optimise accuracy and processing speed.
%Stelios: SVD can be applied on (block-)circulant matrices - so E-RNN is complementatry to our work.

\subsubsection{Run-time methods}
This class of methods exploits techniques to dynamically skip unnecessary computations during the execution of an LSTM. 
In this context, DeltaRNN \cite{fpga_2018} employs a strategy to dynamically avoid computations based on the estimated impact on the output of the network. The skipping criterion is based on the degree of change of each input activation. To effectively implement this technique without significantly dropping the accuracy, the target LSTM has to be trained using the Delta Network scheme \cite{dn_icml_2017}. From a hardware perspective, to overcome the inefficiency of GPUs due to the conditional execution strategy, the DeltaRNN-trained LSTM is mapped on a custom accelerator design which exploits the reconfigurability of FPGAs to efficiently perform the dynamic computations. Nevertheless, despite the run-time computation-skipping, DeltaRNN has requires the target model to be trained using the Delta Network algorithm and hence is limited to settings where the training set is available, while requiring substantial modification of the training scheme and tuning of the hyperparameters. In contrast to this, our method avoids the time overhead and engineering effort of training and parameter tuning and can be directly applied to pre-trained LSTMs.

% A similar approach is proposed in \cite{micro_2018} tailored to mobile-grade GPU architectures. This work employs dynamic row skipping by predicting near-zero results at run time and omitting all preceding computations. Compared to DeltaRNN, this approach does not require modification of the training scheme and hence can be applied directly on pre-trained models post-training. The implementation is optimised to exploit the characteristics of mobile GPUs.

Among the existing designs, reduced arithmetic precision schemes have also been used to obtain gains in terms of performance and power efficiency. ESE \cite{Han_2017} and E-RNN \cite{ernn_2019} employ 12-bit fixed-point precision for both weights and activations. However, to avoid the severe degradation of accuracy due to the limited numerical precision, a fine-tuning step is required by means of additional training iterations. Alternatively, DeltaRNN \cite{fpga_2018} avoids fine-tuning and employs a 16-bit fixed-point representation. Nonetheless, DeltaRNN's quantisation is not network-agnostic, but hand-tuned to minimise the accuracy losses of the target network. In our work, 32-bit single-precision floating-point format is used to avoid the need for fine-tuning and limit the sources of quality-of-result degradation to our approximate computing techniques. Nevertheless, our method is orthogonal and independent of employed numerical precision and thus can be combined with existing quantisation schemes to further boost both performance and power efficiency.

\section{Conclusion}
The deployment of LSTMs in latency-critical applications is a challenging task due to their high computational requirements. In this paper, an iterative approximate computing method together with an FPGA-based architecture are introduced combining model pruning with computation restructuring to make approximate, but informed LSTM predictions in time-constrained environments. In a self-driving car scenario, the proposed system demonstrates significant improvements in accuracy for every given computation time budget compared to a baseline that follows conventional implementations.

 It is noteworthy the proposed approximation methodology effectively reduces the %computational
 workload required to achieve a desired quality of result for a given model and therefore it can be decoupled from the proposed custom-hardware implementation and adapted for deployment on other computing platforms with variable performance gains. Future work encompasses an investigation of ways to adapt the proposed methodology for efficient deployment on other platforms.

\vspace{-3mm}
\section*{About the Authors}
\small{
\textbf{Alexandros Kouris} (a.kouris16@imperial.ac.uk) is a Ph.D. candidate with the Electrical and Electronic Engineering Department of Imperial College London, UK.\\
\textbf{Stylianos I. Venieris} (stylianos.venieris10@imperial.ac.uk) received the PhD degree from Imperial College London and is currently a researcher at Samsung AI Center Cambridge.\\
\textbf{Michail Rizakis} (michail.rizakis14@imperial.ac.uk) is a graduate of the Electrical and Electronic Engineering Department of Imperial College London, UK.\\
\textbf{Christos-Savvas Bouganis} (ccb98@ic.ac.uk) is a Reader in Intelligent Digital Systems in the Electrical and Electronic Engineering Department of Imperial College London and Senior Member of IEEE.
}
\vspace{-3mm}
{
\section*{Acknowledgement}
\small{The support of the EPSRC Centre for Doctoral Training in High Performance Embedded and Distributed Systems (HiPEDS, Grant Reference EP/L016796/1) is gratefully acknowledged. This work is also supported by EPSRC grant 1507723. }
}
\vspace{-2mm}

%\section*{References}
\bibliographystyle{IEEEtran}
\bibliography{Bibliography}

\end{document}